\documentclass{elsart}
\usepackage{amssymb}
\usepackage{epsfig}

\newcommand{\be}{\begin{equation}}
\newcommand{\ee}{\end{equation}}
\newcommand{\bea}{\begin{eqnarray}}
\newcommand{\eea}{\end{eqnarray}}

\newcommand{\eps}{$\epsilon$ }

\begin{document}

\begin{frontmatter}

\title{Generalized dynamical entropies in weakly chaotic systems}
\author{Henk van Beijeren} 
\address{Institute for Theoretical Physics\\
Utrecht University\\ 
Leuvenlaan 4, 3584 CE Utrecht, The Netherlands}

\date{\today}

\begin{abstract}
A large class of technically non-chaotic systems, involving scatterings of light particles by flat surfaces with sharp boundaries, is nonetheless characterized by complex random looking motion in phase space. For these systems one may define a generalized, Tsallis type dynamical entropy that increases linearly with time. It characterizes a maximal gain of information about the system that increases as a power of time. However, this entropy cannot be chosen independently from the choice of coarse graining lengths and it assigns positive dynamical entropies also to fully integrable systems. By considering these dependencies in detail one usually will be able to distinguish weakly chaotic from fully integrable systems.
\end{abstract}
\end{frontmatter}

\maketitle
\section{Introduction}

Dynamical entropies measure rates at which information may be gained by 
observing a system of interest subjected to some law of temporal evolution.
For deterministic dynamics one considers the time evolution of an initial 
ensemble of systems 
distributed uniformly inside a small compact box of diameter $\epsilon$ 
(typically a ball or a cube) within the phase space characterizing the possible 
states of the system. One then asks how many boxes of size $\epsilon$ are at 
least needed to cover the image of the initial box after a time $t$ has elapsed.
If, in the limit $\epsilon \to 0$, the natural logarithm of this number increases as
$h_{top}t$ asymptotically for $t \to \infty$, the quantity $h_{top}$ is called 
the topological entropy. In physical applications usually the Kolmogorov-Sinai
entropy, designated as $h_{KS}$ is more relevant. This is defined in similar 
way, but the covering set is not required to contain the full image of the 
initial box; it only has to cover this up to a subset of measure $\mu(t)$ 
satisfying $\lim_{t\to\infty} \mu(t)=0$. To interpret this, assume \eps is the
minimal resolution with which one may distinguish phase space points by physical 
observation. Then by repeating the observations at regular time intervals the 
number of distinguishable initial boxes increases as $\exp h_{KS}t$ for a 
typical large but not exhaustive number of observed systems and it increases as 
$\exp h_{top}t$ for an exhaustive search in which also rare events are weighted 
properly for the full time range of observation.

For chaotic dynamical systems that are closed (no points escape from phase 
space) Pesin's theorem equates the KS-entropy to the sum of the positive 
Lyapunov exponents. Intuitively this result is not hard to understand. Each 
positive Lyapunov exponent $\lambda_i$ corresponds to an independent direction 
in phase space with a smooth expansion by a factor of roughly $\exp\lambda_i t$ 
over a time $t$. So the overall expansion factor, and thereby the number of 
boxes needed to cover the image of the initial box, increases as $\exp\sum_i 
\lambda_i t$. A priori one could expect though, that, at least in part, this may 
be offset by shrinkage in the contracting directions. However, this contraction 
typically takes place towards a fractal set with a striated structure on 
increasingly finer length scales. Once the typical length scale of separation 
between striation sheets falls below the box size \eps  the number of boxes
required to cover the set does not decrease any more. For comparison: consider 
the number of intervals of length \eps needed to cover successive approximations 
of the middle-third Cantor set. Once the length of the subintervals that are 
erased at the next iteration falls below \eps the number of covering intervals 
essentially does not change any more. The difference between topological and KS-entropy may be understood again by observing that on subsets of phase space of measure zero the Lyapunov exponents may differ from the "bulk" Lyapunov exponents, which for an ergodic system are constant almost everywhere.

There exist dynamical systems that are not chaotic, in the ususal sense of 
having at least one positive Lyapunov exponent, but still show a fairly rapid 
decay to equilibrium, related to motion in phase space that appears very random.
Simple examples of these are {\em wind tree models}, in which light particles
without any mutual interaction move elastically among fixed polygonal 
scatterers. Lepri et al.\cite{rondoni} numerically studied such a model 
consisting of diamond shaped scatterers with corner angles that are irrational 
fractions of $\pi$ on a triangular lattice with bounded free paths between 
collisions. They found this model equilibrates rapidly on the projection to the 
unit cell in space times the unit sphere for the velocities, and on large time 
and length scales it is described by normal diffusion. Dettmann et 
al.\cite{dettmann} looked at Ehrenfests original wind tree model\cite{ehr}, in 
which the scatterers are randomly distributed parallel squares and the 
velocities of the light particles always point along the diagonals of the 
squares. In this case the velocities equilibrate among the four allowed 
directions and, in finite volume the density
of a cloud of light particles becomes uniform. Again, on large time and length 
scales the dynamics becomes diffusive, provided the scatterers are not allowed 
to overlap each other\cite{haugecohen,woodllado,vbh}. Another variation, studied 
by Li et al.\cite{wanghu}, consists of a channel lined by triangular scatterers,
through which light particles move, giving rise again to diffusive motion on 
large scales. Other non-chaotic systems leading to macroscopically stochastic 
looking motion are mixtures of point particles of different masses in one 
dimension. These, for $N$ particles, may be viewed as a type of wind tree model 
on a $2N$-dimensional phase space bounded by flat hyperplanes. The  
intersections between these hyperplanes take the role of the corners; "hitting 
the other side of a corner" translates to an interchange in the sequence of two 
subsequent collisions. Livi et al.\cite{livi} and Grassberger et 
al.\cite{grassberger}
found that such systems show anomalous hydrodynamics, which is typical for 
one-dimensional systems. As a last example I would like to mention systems of 
many moving parallel oriented polygons (in 2d) or polyhedra (in 3d), which in 
collisions exchange the velocity components normal to the plane of collision.
These systems too may be viewed as wind tree like models in a high dimensional 
phase space.
No doubt they will show normal hydrodynamics (up to the usual nonlinearities
in 2d, see e.g.\ \cite{vdh}) although they are obviously non-chaotic. But to my knowledge
no detailed investigations have been done for such systems so far.

For the systems described above, according to Pesin's theorem the KS-entropy equals zero. Yet, 
it is clear that information may be gained by repeated observation, only the 
number of distinguishable initial boxes will not increase exponentially with time, but 
rather as a power law. Zaslavsky and Edelman refer to this situation as
pseudochaos\cite{zaslav}. I will use the term "weak chaos" in the sequel, without giving a precise 
definition. For quantifying the power law increase of the information on the initial 
distribution Tsallis' formalism seems 
suitable, as it is precisely meant to define an entropy for systems in which the 
countable number of states (here distinguishable initial boxes) does not increase exponentially 
with system size (here time) but rather as a power law. But there are some 
subtleties here to deal with. As we will see, the dynamical entropy as defined usually, will become dependent on the coarse graining lengths in the system. It has to be multiplied by powers of these to allow for taking the limit $\epsilon \to 0$. In addition, even very regular systems, such as 
an ideal gas in a periodic box, or a weakly anharmonic oscillator, will turn out to give 
rise to a positive dynamic Tsallis entropy, thereby excluding this property by itself as 
an indicator of weak chaos. Keeping track of the power law by which the 
information increases may help to recognize weak chaos. Even this requires a lot 
of care, because the powers of time resulting from integrable degrees of 
freedom may vary, depending on the type of motion they represent.

In the next sections we will consider dynamical Tsallis entropies and look at 
wind tree models in detail. In the discussion
the feasibility of using a dynamical Tsallis entropy to characterize weak chaos 
will be discussed further.

\section {Dynamical Tsallis entropies}\label{sec:de}
Supose the phase space of a class of dynamical systems is divided into subsets
$C_i$ such that the probability of finding a system in $C_i$ equals $p_i$. Then 
the Tsallis entropies are defined through.
\be
S_q=\frac{1-\sum_ip_i^q}{q-1.}
\ee
Here $q$ is a free parameter the choice of which should be dictated by the 
properties of the system considered. In the limit $q\to 1$ the Tsallis entropy
reduces to the usual entropy.

In the case of dynamical systems evolving in time from the interior of an
initial set of diameter \eps one may choose for the $C_i$ a covering set of 
non-intersecting subsets of diameter \eps and define $p_i(t)$ as the
fraction of initial phase points that end up inside $C_i$ after the 
time evolution over time $t$. The number of $C_i$ with non-vanishing $p_i(t)$ will increase as a power of $t$. Now let us suppose that on this subset $p_i(t)$
for large $t$ behaves as
\be
p_i(t)=\frac{w_i}{Ct^\alpha},
\label{psubi}
\ee
with $w_i$ distributed according to some probability distribution with a mean 
value of unity. Then, in order for $S_q(t)$ to increase proportionally to $t$ 
one has to 
choose $q=1-1/\alpha$. In this case the entropy behaves as
\be
S_{1-1/\alpha}(t)=\alpha C^{1/\alpha}<w_i^{(1-1/\alpha)}>,
\ee
with the brackets indicating an average over the distribution of $w_i$. In case 
this is an exponential distribution, $p(w)=\exp -w$, one finds
$<w_i^{(1-1/\alpha)}>=\Gamma(2-1/\alpha)$.

A further remark is required here. As time increases
one should choose increasingly smaller values of $\epsilon$ in order to avoid $C_i$'s containing parts of the image of several initial boxes. For fixed $\epsilon$ this will always occur after sufficiently long time. In principle one should take a limit $\epsilon \to 0$ followed by the limit $t \to \infty$. We will see however that the generalized dynamical entropies one would like to define are dependent on $\epsilon$, so one cannot take these limits in a straightforward way. In fact, as we will see, it can only be done after redefining the dynamical entropies in an appropriate way.

\section{Wind tree models}
Here I will consider a wind tree model in two dimensions with square
scatterers of side length $a$ that are randomly oriented and distributed 
randomly over a rectangle with periodic boundary conditions.
\begin{figure}
\centerline{\epsfig{file=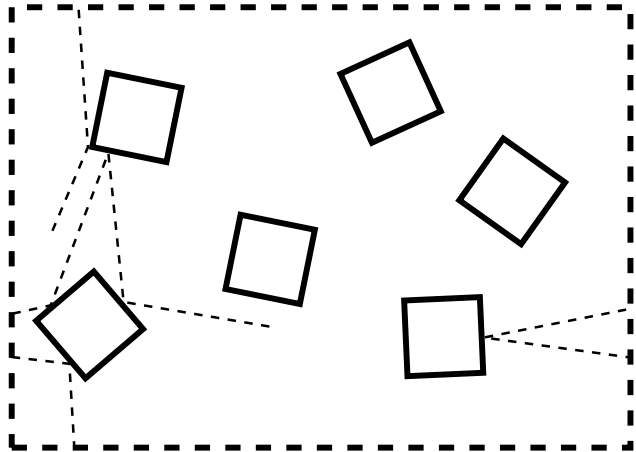,height=6cm}}
\caption{A configuration of the 
wind tree model with randomly oriented square scatterers. The dashed line represents a posible light-particle trajectory (note the periodic boundary conditions)}
\end{figure}
The scatterers are not allowed to overlap each other. I will assume that the density 
$\rho$ of scatterers is small, that is $\rho a^2\ll 1$, so subsequent collisions 
of a light particle are to a very good approximation uncorrelated. For the 
subsets $C_i$ of the phase space let us choose small rectangular blocks of side
lengths $\epsilon_r$ in position space and $\epsilon_v$ in velocity space. Over 
a time $t$ the dynamics stretches out these blocks into parallellopipida. For large $t$, 
in the direction parallel to the velocity these obtain an extension of length 
$\epsilon_v t$. 
In the direction orthogonal to the 
velocity the front width covered by the parallellopipidum also grows as $\epsilon_v t$. 
In addition the block is split 
into two subblocks, as illustrated in Fig.\ 2, as soon as its front hits a 
corner of one of the scatterers.
\begin{figure}
\centerline{\epsfig{file=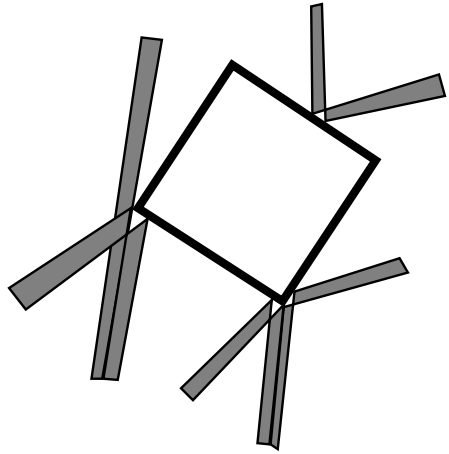,height=6cm}}
\caption{When a block of nonzero density in phase space hits a corner of a scatterer it is split into two subblocks. If it just hits a side it is reflected in its entirety.}
\end{figure}
 Next, when one of the two subblocks hits 
another corner it is split up likewise and so on. The total front width of all 
the propagating subblocks increases with time as $\epsilon_v t$, independent of 
the number of splittings that have taken place. Therefore the average number of 
splittings during a long time $t$ is given by
\be
N(t)=\frac 3 2 \rho v\epsilon_v t^2.
\ee
Here $v$ is the average speed of a light particle. The average area swept out by 
the total front of all the subblocks is $v\epsilon_v t^2/2$ and this is 
multiplied by $3\rho$ to obtain the average number of visible corners in this 
area. The average number of resulting subblocks is almost equal to $N(t)$. To 
obtain the number of subsets $C(i)$ needed to cover all of the subblocks at $t$
one has to multiply $N(t)$ by $\epsilon_v t/\epsilon_r$, the number of subsets 
needed to cover the extension in the parallel direction. So in Eq.\ (\ref{psubi}) 
$\alpha=3$ and $C=3 \rho v\epsilon_v^2/2\epsilon_r$.
The positions of the corners with which the collisions take place follow a 
Poisson distribution along the front, hence the distribution of $w_i$ is 
an exponential one indeed. Using the results of the preceding section we find
\be
S_{2/3}(t)=\alpha t\left(\frac {3 \rho v\epsilon_v^2}{2\epsilon_r}\right)^{2/3}\,\Gamma\left(\frac 5 3\right).
\label{S}
\ee

\section{Discussion}
The first thing that draws the attention in Eq.\ (\ref{S}) is the dependence of the entropy on the coarse graining lengths $\epsilon_v$ and $\epsilon_r$. This is a striking difference from the result found by Borges et al.\cite{borges} for the logistic map at the onset of chaos. Here too the number of boxes increases 
algebraically with time, but this is due to a power law separation between 
infinitesimally close trajectories in phase space. In the wind tree models 
infinitesimally close trajectories typically just separate linearly with time,
due to the infinitesimal velocity difference, but the interesting power law 
behavior in the time dependence of the number of covering sets needed is caused by the corner induced splitting of trajectories at 
arbitrarily small, but always finite distances from each other. The same is true for all the other examples mentioned in the introduction. As mentioned at the end Sec. \ref{sec:de} a sharp definition of dynamical entropies involves taking a limit $\epsilon \to 0$ followed by the limit $t \to \infty$. Obviously this makes no sense for the entropy obtained in Eq. (\ref{S}). The limit  $\epsilon \to 0$ is not well-defined. Apparently the only way to repair this is by defining
still another dynamical entropy as
\be
\tilde{S}_q=\left(\frac{\epsilon_r}{\epsilon_v^2}\right)^q S_q.
\label{salt}
\ee
An alternative way of obtaining this is by requiring $\epsilon_r=\epsilon_v^2$.
For wind tree like models in $d$ dimensions the prefactor $\epsilon_r/\epsilon_v^2$ should be raised to the power $d-1$. This solution is not very elegant, e.g. 
there seems to be no obvious choice for making 
the $\epsilon$'s dimensionless. But at least it seems to give a proper limiting behavior for $\epsilon \to 0$ for all wind tree like models.

A second point of concern is the entropy of an ideal gas in a periodic box. One easily sees that, with 
the same type of covering sets as for the wind tree model, 
the number of sets needed to cover the time evolved initial 
set increases in $d$ dimensions as $(\epsilon_v t/\epsilon_r)^d$. So for $q=1-1/d$ one now has a dynamical entropy that increases linearly with time. The same occurs for other 
integrable systems in which the periods of the angle variables depend on the 
values of the action variables. This is usually the case, with harmonic 
oscillators being the notable exception. For $d$-dimensional wind tree models
the power of $t$ by which the number of needed covering sets increases with time is 
$2d-1$. So it looks like one could use the condition $q>1-1/d$ as a criterion 
for weak chaos in a $d$-dimensional hamiltonian system. Here too care is needed.
For example in a three-dimensional system that is a wind tree model in the 
$x-y$ plane but in addition performs a harmonic oscillation in the 
$z$-direction,
one has $q=1/3$, just as for a three-dimensional ideal gas. A distinction between 
the two cases may still be found in the dependence of $S_{2/3}$ on $\epsilon_r$ 
and $\epsilon_v$. The powers of $\epsilon_v$ and $\epsilon_r$ by which one has to multiply $S_q$ in order to obtain a $\tilde{S}$ with a proper limit for 
$\epsilon \to 0$ obviously are different for integrable systems of various types, for wind tree like models and for mixed systems.

It is not hard constructing generalizations of the wind tree model to higher 
dimensions. One just has to replace the scatterers by polyhedra of appropriate
dimension. And indeed for each direction orthogonal to the velocity one picks up 
a factor $t^2$ in the growth factor of $N(t)$.  Similar observations can be made 
for the moving polygons.

To conclude: Dynamical Tsallis entropies may be useful for characterizing
weakly chaotic systems in which the information on initial conditions increases 
as a power law rather than an exponential. For wind tree like systems however, 
the precise value of this entropy depends on the choice of the coarse-graining 
lengths, which basically sets the smallest scales on which one may distinguish 
different systems in a single observation. To obtain a dynamical entropy with a proper limit when these coarse graining lengths are sent to zero, one first has to multiply the "straightforward" dynamical entropy with appropriate powers of the coarse graining lengths. Further, the generalized entropies typically assume non-zero values also for integrable systems. In 
many cases the $q$-values needed to find an entropy that increases linearly with 
time will allow one to distinguish wind tree like systems from integrable ones, 
but this too is not always the case. If not, the dependence on the coarse 
graining lengths may be used for a further distinction.

HvB acknowledges support by the Mathematical physics program of
FOM and NWO/GBE.

\bibliographystyle{prsty}

\begin{thebibliography}{12}
\bibitem{rondoni} S.\ Lepri, L.\ Rondoni and G.\ Benettin, J.\ Stat.\ Phys.\
	{\bf 99} (2000) 857
\bibitem{ehr} P.\ Ehrenfest, Collected scientific papers (North Holland, 
	Amsterdam 1959)
\bibitem{dettmann}C.\ P.\ Dettmann, E.\ G.\ D.\ Cohen and H.\ van Beijeren, 	
Nature {\bf 401} (1999) 875; C.\ P.\ Dettmann and E.\ G.\ D.\ Cohen, J.\ 	
	Stat.\ Phys.\ 103 (2001) 589
\bibitem{haugecohen} E.\ H.\ Hauge and  E.\ G.\ D.\ Cohen, J.\ Math.\ Phys.\ 
{\bf 10} (1969) 397
\bibitem{woodllado}  W.\ W.\ Wood and F.\ Lado, J.\ Comp.\ Phys.\ 
	{\bf 7} (1971) 528
\bibitem{vbh} H.\ van Beijeren and E.\ H.\ Hauge, Phys.\ Lett.\ 
	{\bf 39A} (1972) 397
\bibitem{wanghu} B.\ Li, L.\ Wang and B. Hu, Phys.\ Rev.\ Lett.\ {\bf 88} (2002) 
223901 
\bibitem{livi} R.\ Livi, A.\ Politi and S.\ Ruffo, J.\ Phys.\ A: Math.\ Gen.\
	{\bf 19} (1986) 2033 
\bibitem{grassberger} P.\ Grassberger, W.\ Nadler and L.\ Yang, Phys.\ Rev.\ 
	Lett.\ {\bf 89} (2002) 180601
\bibitem{vdh} M.\ A.\ van der Hoef and D.\ Frenkel,
   	 Phys.\ Rev.\ {\bf A 41} (1990) 4277
\bibitem{zaslav} G.\ M.\ Zaslavsky and M.\ Edelman, nlin.CD/0112033
\bibitem{borges} E.\ P.\ Borges, C.\ Tsallis, G.\ F.\ J.\ A\~na\~nos
	and P.\ M.\ de Oliveira, Phys.\ Rev.\ Lett.\ {\bf 89} (2002) 254103

\end{thebibliography}

\end{document}